# Spatial dependency between task positive and task negative networks

Robert Leech[1]*; Gregory Scott[1]; Robin Carhart-Harris[1]; Federico Turkheimer[2]; Simon D. Taylor Robinson[3]; David J. Sharp[1]

[1]Division of Brain Sciences, Hammersmith Hospital, Imperial College London, Du Cane Road, London, W12 0NN, United Kingdom, [2]Department of Neuroimaging, Institute of Psychiatry, King's College London, De Crespigny Park, London SE5 8AF, [3]Department of Medicine,10th Floor QEQM Wing, St Mary's Hospital Campus, Imperial College London, South Wharf Street, London W2 1NY, United Kingdom.

## Abstract

Functional neuroimaging reveals both relative increases (task-positive) and decreases (task-negative) in neural activation with many tasks. There are strong spatial similarities between many frequently reported task-negative brain networks, which are often termed the default mode network. The default mode network is typically assumed to be a spatially-fixed network; however, when defined by task-induced deactivation, its spatial distribution it varies depending on what specific task is being performed. Many studies have revealed a strong *temporal* relationship between task-positive and task-negative networks that are important for efficient cognitive functioning and here. Here, using data from four different cognitive tasks taken from two independent datasets, we test the hypothesis that there is also a fundamental *spatial* relationship between them. Specifically, it is hypothesized that the distance between task positive and negative-voxels is preserved despite different spatial patterns of activation and deactivation being evoked by different cognitive tasks. Here, we show that there is lower variability in the distance between task-positive and task-negative voxels across four different sensory, motor and cognitive tasks than would be expected by chance - implying that deactivation patterns are spatially dependent on activation patterns (and vice versa) and that both are modulated by specific task demands. We propose that this spatial relationship may be the macroscopic analogue of microscopic neuronal organization reported in sensory cortical systems, and we speculate why this spatial organization may be important for efficient sensorimotor and cognitive functioning.

# Introduction

Functional neuroimaging studies often typically show coupled increases, as well as decreases in regional metabolism and blood flow. The pattern of relative increases in brain activity is highly variable, and depends on the specific nature of task demands. In contrast, relative decreases in activation appear at first glance to have similar spatial pattern, regardless of the nature of the task. Relative deactivation is often observed in the lateral and medial inferior parietal lobes (including the posterior cingulate cortex) and the ventromedial prefrontal cortex. Collectively these regions are termed default mode network (DMN) (1, 2). This pattern of regional deactivation is broadly preserved in homologous regions in non-human primate species (3) and has also been observed in rodents (4).

The existence of a temporal relationship (often, an anti-correlation) between some variant of the DMN and some task-positive networks is thought to be important for efficient cognitive functioning (5). For example, many cognitive tasks (e.g., a two-back working memory task) evoke increases in activity within fronto-parietal networks associated with cognitive control and attention which are tightly coupled with decreases in activity within the DMN (6, 7) (although see Spreng, 2012, for a discussion of situations where the DMN and cognitive control networks vary together). Similarly, the strength of the temporal anti-correlation between the DMN and dorsal attention network/FPCN is associated with behavioral performance (8). However, depending on the task or intervention, the temporal relationship between the DMN and executive networks is not always an anti-correlation (9-11).

Although a deactivation pattern roughly consistent with a canonical DMN is reported for many different tasks, the precise pattern of regional deactivation varies with the specific task requirements (7, 12, 13). This may imply that the DMN is not a spatially fixed entity but rather a malleable phenomenon that varies in a predictable way depending on the specific task demands. We hypothesize that the specific spatial distribution of an evoked deactivation pattern is, to a significant degree, molded by the activation pattern evoked by a specific task. In a similar vein, a reactive spatial relationship between positive and negative networks, the hypothesized spatial relationship might imply some level of homeostatic dependency between these generic brain systems.

Here, we test the hypothesis that there exists a spatial dependency between task positive and negative networks. In what follows, we use the terms task-positive (TP) and task-negative (TN) to mean the specific pattern of relative activation and relative deactivation evoked by a specific task, without restricting this to any canonical functional brain networks (e.g., TN does not necessarily correspond to the DMN). We use data from four different cognitive tasks taken from two independent datasets to test the spatial relationship between the networks. A detailed description of the analytical method is presented in Figure 1. We predict that if there is a spatially dependent relationship between TP and TN networks, then the distance between TP and TN voxels should be relatively preserved across trials (i.e., lower variability in distance between TP and TN voxels than would be predicted by chance) (top row of Figure 1). To test this, we quantified the shortest distance between TP and TN voxels, and then assessed the variability of these distances using both the Gini coefficient and the coefficient of variation. We then repeated this calculation for TP and TN voxels from different tasks, which allowed us to test whether the variance in the spatial distance between TP and TN voxels was lower within a task (top row of figure 1) than between tasks (bottom row of Figure 1). If there is no task-specific modulation of the TP and TN networks spatial relationship, then the distributions of distances should be, on average, the same whether the TP and TN patterns are evoked by the same or different tasks. On the other hand, if there is a task-dependent modulation of the spatial relationship between the TP and TN, then there should be lower variability for the within-task TP to TN distance (because both networks are being 'pulled' into a specific configuration to serve a specific function) than when comparing across different tasks (because different tasks will 'pull' the networks in different ways - depending on what specific functions need to be served).

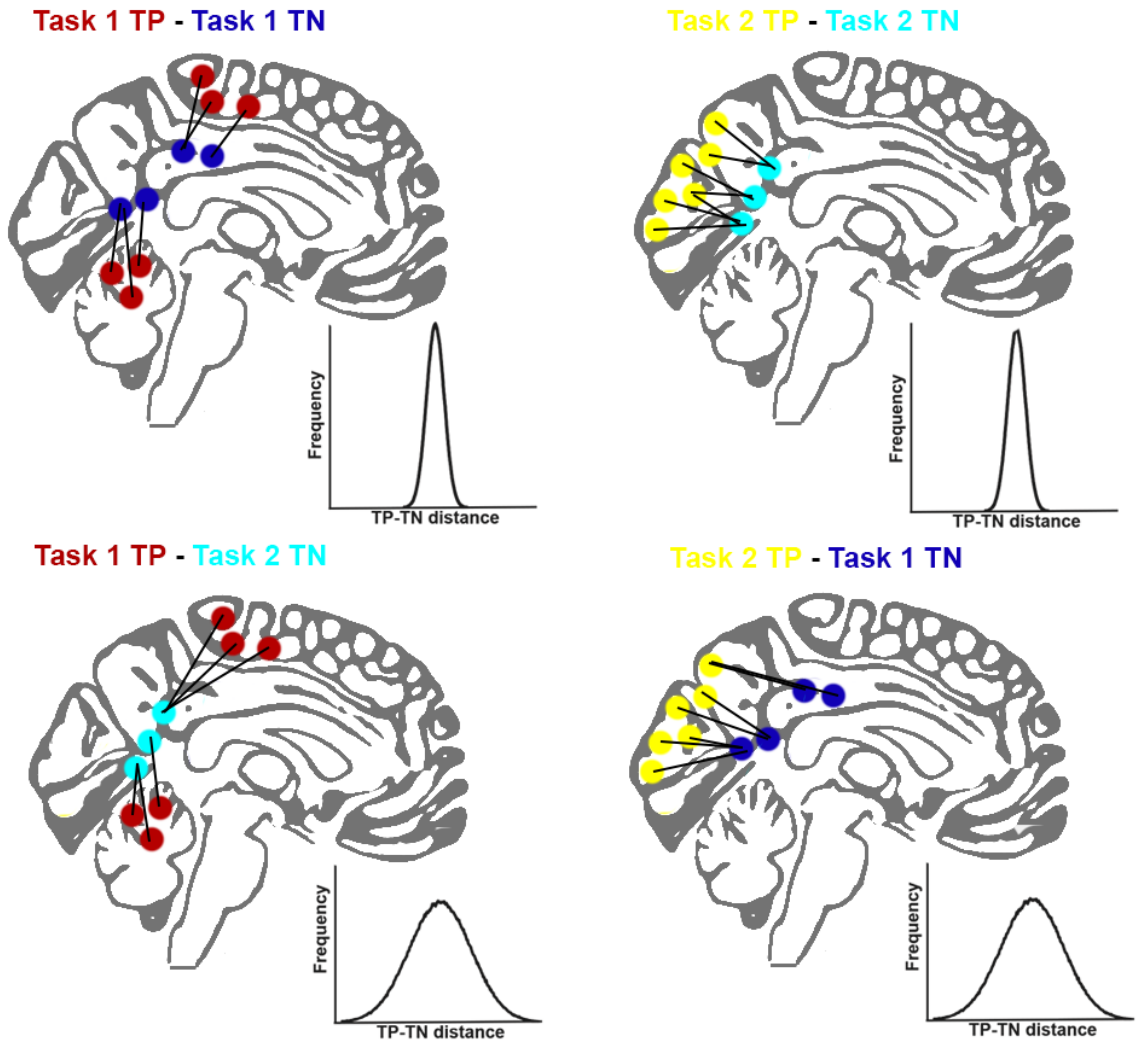

*Figure 1: Schematic showing task positive (TP) and task negative activations (TN). The illustration schematically shows how the minimum distance between TP and TN voxels (small colored circles) was calculated on two hypothetical tasks. This is done for the TP and TN voxels from the same task and for TP voxels from one task and TN voxels from the other task. If the variable 'task' did not modulate the spatial relationship between the TP and TN networks, we would expect uniform distributions (i.e., represented by the width and height of the distribution) in each of the four cases. However, the schematic shows how variance in TP-TN distance increases between different-tasks (top versus bottom) relative to within the same task (left versus right). .*

# Methods

*Participants*

All participants gave written consent, were checked for contraindications to MRI scanning and had no history of significant neurological or psychiatric illness. The Hammersmith, Queen Charlotte's and Chelsea research ethics committee awarded ethical approval for the study. Dataset 1 involved 21 neurologically healthy adult subjects (mean (SD) age 44 (9) years, 11 female). These subjects were scanned while alternating between viewing a static chequerboard, performing a simple finger tapping task or the rest condition with no explicit task. Dataset 2 involved 18 neurologically healthy adult subjects (mean (SD) age 29 (6), 9 female) who performed a working memory task and a cued autobiographical memory task.

*MR Imaging acquisition and analysis*

Images on all participants were acquired using a 3 Tesla MR scanner (Intera™, Philips Medical Systems, Best, Netherlands). We acquired a high-resolution $T_1$-weighted sequence and an 8-channel head coil.

Dataset 1: A single run was acquired while subjects performed simple visual or motor tasks. $T_2$*-weighted gradient echo-planar images (EPI) were collected with whole-brain coverage, with the following parameters: TR=3s; TE=30 ms; alpha=90°, 2.2 X 2.2 axial slices, slice thickness 2.75 mm. Quadratic shim gradients were applied to correct magnetic field inhomogeneity. Subjects observed a flashing chequerboard or tapped their right or left finger on their thigh. The experiment used a blocked design with visual stimulation, finger tapping or a baseline rest condition, alternating in blocks of 24 s. There were 150 acquisitions, lasting 7.5 min.

*Dataset 2:* Data were acquired while subjects performed two separate runs: i) a 2-back working memory task; ii) an autobiographical memory task. $T_2$*-weighted gradient echo planar images were collected with whole-brain coverage, with the following parameters: repetition time, 2 s; echo time, 30 ms; alpha = 90°; 31 slices; slice thickness 3.25 mm; interslice gap of 0.75mm; acquisition in ascending order (resolution: 2.19, 2.19, 4.0 mm). Quadratic shim gradients were used to correct for magnetic field inhomogeneities within the brain. The 2-back working memory task was a blocked design. There were five 36 second blocks of task, interspersed with 10 second rest blocks. The task periods consisted of the 2-back task visually presented with pictorial stimuli Preparation for the autobiographical memory task involved subjects providing a list of written memory cues (e.g., "Remember the engagement party") prior to scanning. In the scanner, these cues were visually presented and the subjects were asked to think about the autobiographical memory cued. As with the 2-back task, five 36 second blocks of the autobiographical memory task were interspersed with 10 second rest blocks.

Whole-brain fMRI data were analyzed individually with standard general linear models analysis tools from the FSL library (FEAT version 5.98). For all runs from all subjects, image pre-processing involved realignment of EPI images to remove the effects of motion between scans, spatial smoothing using a 6 mm full-width half-maximum Gaussian kernel, pre-whitening using FMRIB's improved linear model (FILM) and temporal high-pass filtering using a cut-off frequency of 1/50 Hz to correct for baseline drifts in the signal. FMRIB's Linear Image Registration Tool was used to register EPI functional datasets into standard MNI space using the participant's individual high-resolution anatomic images. FMRI data of each individual subject were analyzed using voxel-wise time series analysis within the framework of the General Linear Model (GLM). To this end, a design matrix was generated with a synthetic hemodynamic response function and its first temporal derivative. The timecourses from motion parameters were also included to attempt to account for noise related to head motion. For dataset 1, timecourses of visual stimulation and left and right finger tapping were modeled in the design matrix for each subject, with rest as an implicit baseline. For dataset 2, each run was analyzed separately, with a simple design matrix modeling the blocks when tasks were being performed versus the implicit rest baseline (either 2-back or autobiographical memory or task).

T-statistic images for each individual for each task were calculated. These statistical maps were then thresholded at a range of values (both positive to derive task positive maps and negative values for task negative maps), binarized, resampled into 4mm x 4mm x 4mm voxel resolution and then used to assess the spatial relationship between TP and TN (see below) voxels. To illustrative the approximate distributions of TP and TN voxels used in the spatial variability analyses, group maps were created from the individual t-stat images. Each subject's t-stat maps was thresholded at $t>2$, binarized and subsequently combined across subjects. The percent overlap maps for TP and TN voxels are presented below. Voxelwise higher-level group GLM results, corrected for multiple comparisons are not presented, since our hypothesis does not involve asking whether any specific voxel is active at the group level or not.

### *Assessing the spatial relationship between TP and TN*

Figure 1 illustrates the general approach to measuring the spatial relationship between the TP and TN. To do this, we estimated the minimum Euclidean distance in number of voxels from each TP voxel to the nearest TN voxel. The distribution of minimum distances was then used to calculate different measures of variability: the Gini coefficient and the coefficient of variation. The Gini coefficient is a standardized measure of the equality of a frequency distribution (with a value of 0 meaning that there is no variability and a value of 1 meaning maximal variability) and the coefficient of variation is simply the standard deviation scaled by the mean distance. Both are standardized measures of the variability of a distribution that are unaffected by the mean value of the distribution. Standardization was done because there are different numbers of voxels activated or deactivated in different tasks (e.g., because of differences in the sensitivity of the tasks, regional differences in neural signal and non-neural noise sources). These different numbers of voxels could alter the mean distances between the TP and TN, and as a result affect non-standardized measures of variability. Below, we report results with the Gini coefficient; however, the coefficient of variation provided qualitatively the same pattern of results.

For each dataset there were two tasks, resulting in four possible TP to TN distributions of minimal distance (task 1 TP to TN; task 1 TP to task 2 TN; task 2 TP to task 2 TN; and task 2 TP to task 1 TN: see Figure 1). These variability measures were subsequently statistically compared in a general linear model, with intra/inter-task and task-type as repeated measures factors.

To calculate distances between TP and TN voxels, the data was thresholded at arbitrary t-values in the range $t=1.7$ (nominally approximately $p=0.05$) to 3 ($p<0.005$; at $t>3$, many subjects had no voxels activated for at either the TP or TN in at least one task). Qualitatively similar results were found across the range of t-values; therefore, we present results at voxel thresholds of $t=2$ and $t=3$. Correcting the voxelwise statistics for multiple comparisons was not appropriate at this stage because the hypothesis was not about the presence of voxel specific activation, but rather the brain-wide distribution of TP to TN distances.

To ensure that differences between intra and inter-task spatial variability were not an artifact of extreme values, we calculated the skewness of each frequency distribution. We repeated the spatial variability analyses, excluding all TP to TN distances at or below a minimum Euclidean distance of two voxels (8mm). The minimum distance analysis was performed to test if the results are consistent even without nearby voxels which are more likely to share vasculature. Results below are presented both including all TP to TN distances and excluding Euclidean distances less than two voxels.

## Results

**Task positive and task negative regions at the group level**

The pattern of relative activation (TP) and deactivation (TN) for two tasks (a visual checker board or a finger tapping task) was analyzed across all subjects in the first group (Figure 2). This showed the expected pattern of activation in visual occipital regions (lighter warm colors) for the visual task and sensorimotor parietal, frontal and cerebellar regions for the motor task (darker warm colors). Whereas the different tasks evoked highly distinct spatial patterns for relative activation (TP), the patterns of relative deactivation (TN) were more similar across the two tasks, with some overlapping regions within the lateral parietal and occipital regions. However, despite this broad similarity, the visual (light blue) and motor (dark blue) tasks evoked subtly different patterns of deactivation, with the visual TN voxels being generally more posterior. Therefore, the group data suggest that the different tasks evoke different patterns of activation and deactivation, i.e. spatially different TP and TN networks.

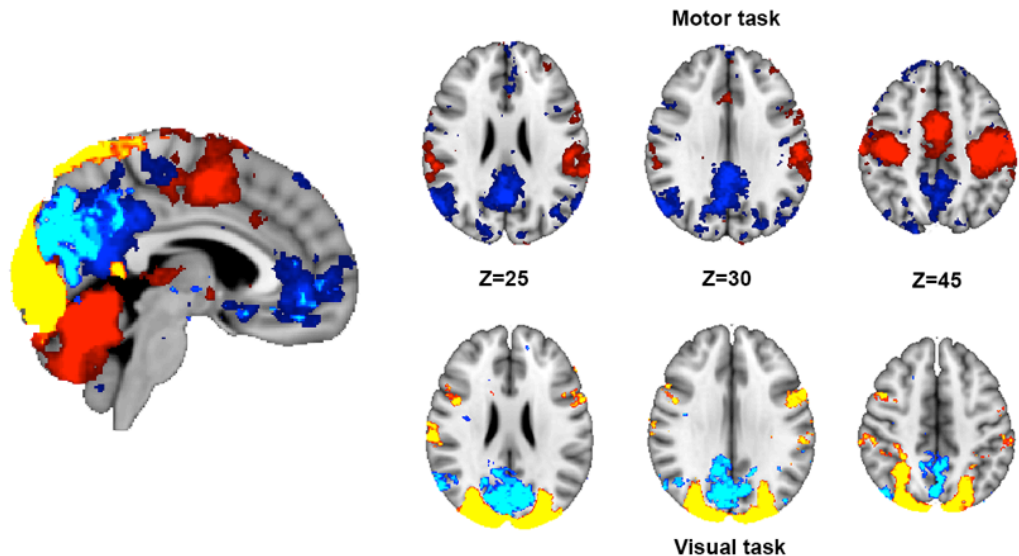

*Figure 2: Voxels that are activated (t-threshold>2 for >20% of subjects) for the motor task are in the red (task positive: task > rest) and voxels deactivated are in dark blue (task negative: rest > task) colors. The visual task is in yellow (task positive) and light blue (task negative). Notice how the deactivation pattern appears to shift so that it is proximal to the activation pattern the two different tasks, thus maintaining a relatively consistent distance between the TP and TN networks.*

### The distance between task-positive and task negative voxels is less variable within a task than across tasks

To assess whether there is task-dependent modulation of the spatial relationship between the TP and TN voxels, the distance between TP and TN voxels was calculated, and the variability of these distances was then compared across and within tasks. Figure 3 shows the variability in the distances (as measured by the Gini coefficient) between visual and motor TP voxels and TN voxels - either across or within tasks, both at the group level (Figure 3A) and for individuals (Figure 3B). A repeated measures general linear model showed that inter-task variability was significantly greater than intra-task variability ($F(1,20)=59.8$, $p<0.001$), consistent with a task-dependent modulation of the TP-TN network spatial relationship. Similar results were also obtained using non-parametric approaches (e.g., Wilcoxon signed-rank tests). This result was observed using a threshold of t>3 to define TP and TN voxels, and the same pattern of results was observed: when the threshold was t>2 (F(1,20)=67.7, p<0.001). Qualitatively similar results were found using a different measure of variability (the coefficient of variability), also at t>2 and t>3.

Since it is logically impossible for within-task TP and TN voxels to overlap, but not for between-task TP and TN voxels to do so, it is theoretically possible that there could be an artificial skew in the within-task conditions towards lower distances than for the between-task data. To rule out this possibility, we examined the skewness of the distances as well as the variability in the distances. We found no significant effect of intra- versus inter-task (F(1,20)=0.029, ns) skewness. As a further check, we also reran the variability analyses, while removing distances that were <2 a minimum Euclidean distance of 2 voxels, apart, eliminating small distances (and making a simple explanation of the results in terms of shared vasculature less likely – see discussion). As with the prior analyses, even while removing small distances, there was a significant main effect of intra- being less variable than inter-task conditions (F(1,20)=33.9, p<0.001).

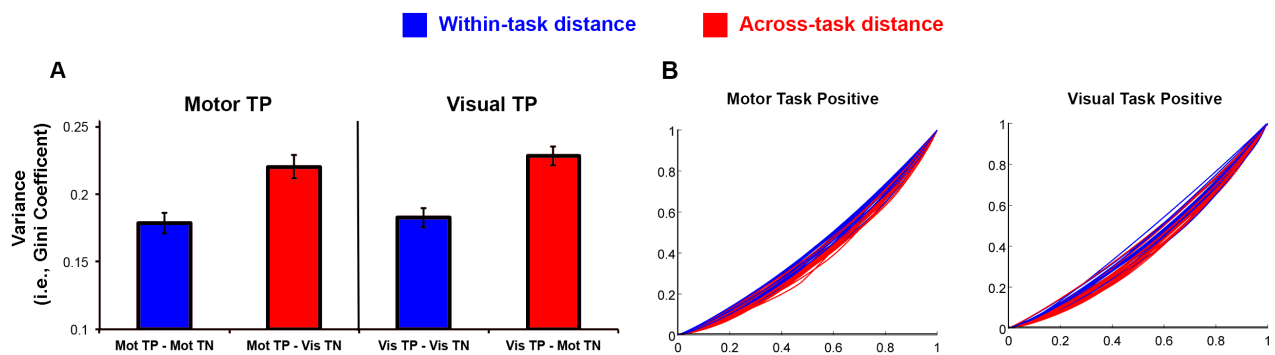

*Figure 3: Variability in the distance between task positive and task negative (measured by the Gini coefficient) voxels for the motor and visual tasks, comparing intra-task versus inter-tasks. 3A plots the average Gini coefficient using a t-threshold >3, before calculating distance. 3B shows individual subjects' Lorenz curves (a graphical representation of the Gini coefficient). For each subject, the distances from smallest to largest are plotted against the cumulative frequency: blue is within task, red is TP-TN distance between tasks (the closer the line is to the 45º line, the lower the variability). The intra-task curves (blue lines) are on average closer to the 45º line than the inter-task curves (red lines) for both visual and motor task positive voxels.*

### Reduced variability in the spatial relationship also occurs in a separate dataset using different tasks

We demonstrated the same result with a second dataset, acquired with different set of subjects performing two different tasks: a working memory 2-back task and a cued autobiographical memory task. Figure 4A shows the pattern of TP and TN voxels across subjects. This shows some overlap with the TP regions for both tasks (e.g., in medial prefrontal regions) but also considerable spatial differences (e.g., in visual, parietal and retrosplenial regions). Similarly, the TN regions were considerably different between the two tasks. Figure 4B shows the variability in TP-TN distances for the different tasks (both intra- and inter-tasks). We observed the same pattern of lower within versus across task variability in TP-TN distance (thresholded at t>3: $F(1,14)=58.2$, $p<0.001$; thresholded at t>2: $F(1,17)=59.8$ $p<0.001$; and with a minimum distance of at least two voxels: $F(1,14)=12.8$, $p<0.005$). As with the previous dataset, there was no evidence of a difference in skewness when comparing within versus across tasks ($F(1,14)=0.3$, ns).

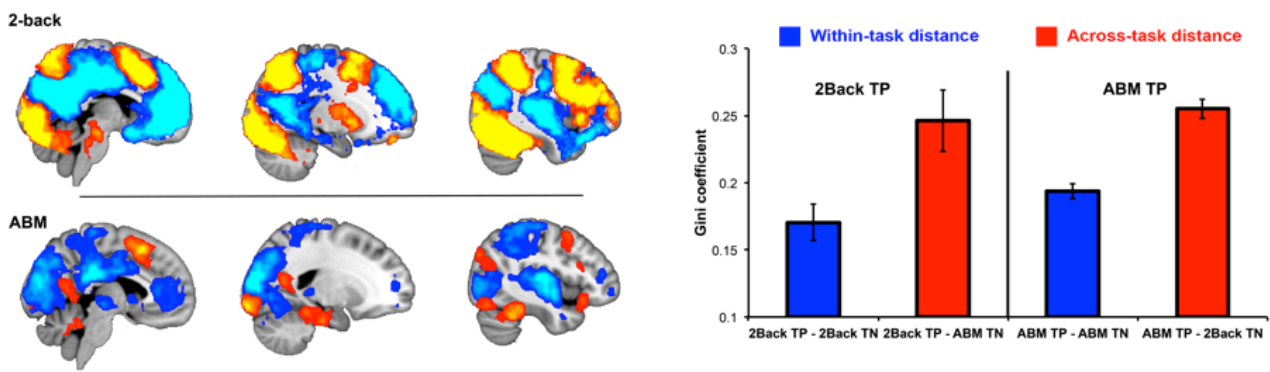

*Figure 4: A second example using two different, cognitive tasks with a different set of subjects. In this case, a 2-back working memory task and an autobiographical memory task. A: TP and TN voxels (t>2, present in >20% of subjects) are presented for the two tasks.*

**Discussion**

Across four different tasks, involving sensory, motor and internally and externally focused processing, the minimum distance between task positive and task negative voxels displayed less variability than would be expected by chance. This result supports the principle that the spatial separation between the TP and TN networks is relatively preserved across different tasks. Moreover,. it suggests that at least some proportion of the TN system varies in a manner that is dependent on the spatial localization of the activation pattern. For example, a visual stimulus activating occipital regions will be accompanied by a deactivation pattern in proximal occipital and medial parietal regions. This relationship may be considered as a spatial analogue of the temporal relationship (e.g., anti-correlations) frequently reported between some task positive and task negative networks (Fox, Snyder et al. 2005 (14, 15). That is, an activation pattern may be counteracted by a deactivation pattern that is spatially proximal to it..

There are a number of possible explanations for the spatial dependency between task positive and task negative systems that has been systematically demonstrated here. One attractive possibility is that due to the extremely high metabolic demands of neural activity (Atwell & Laughlin, 2001), patterns of activation must be counteracted by proximal patterns of deactivation, so to avoid episodes of run-away excitation. Thus, although energetically counteractive, spatially, the task-evoked deactivations appear to be partially attracted to the patterns of activation. This inclination for spatial proximity between the networks may reflect a requirement of the global system to maintain a critical level of activity, perhaps implying that functional brain systems compete for what is a quasi-finite metabolic quota . Similarly, the spatial relationship that we observe between task positive and task negative networks may exist in order to prevent runaway excitation – with the TN network serving as a kind of ballast to resist further activation and stabilize the global system.. It has been argued that the brain operates at or near a critical state (16-18) (which has advantages in terms of optimal information processing and allows the system to explore multiple different states and respond rapidly and efficiently), and that the functional architecture that supports this operates across spatial scales (16, 19). The task negative regions spatially adjacent to task positive regions, may allow the brain to operate at or near this high-energy critical state, without crossing into either super- or sub-critical (e.g., having epileptic seizures).

A related explanation is that the spatial adjacency of competitive systems has benefits for efficient information processing. A model of balanced activity (with neurons with a positive response coupled with nearby neurons with negative responses) has been proposed to operate at the neuronal level within primary visual and auditory cortices (20). Neurons are thought to communicate using efficient coding in which redundant information is discarded (21). Receptive fields of mammalian visual and auditory cortices are spatially organized, operate across a range of spatial scales, and can be understood as banks of Gabor (or wavelet) filters (22-24). These filters remove statistical redundancy and increase independence between neuronal responses to natural stimuli. However natural sounds and images exhibit strong statistical dependencies that cannot be eliminated with linear operations alone.) Schwartz and Simoncelli (2001) proposed a simple and efficient solution to this problem by demonstrating that linear filter banks are optimal for coding as long as the output of each filter is rectified (divided) by a weighted sum of the spatially adjacent (and time/frequency conjoint) filter outputs. Therefore, positive neuronal responses to a sensory input are intrinsically coupled to nearby negative responses. We suggest that this property, found in sensory cortex, may scale-up to describe the system-level patterns of activation observed here at the macroscopic scale. Similarly, this phenomenon may not be limited to just sensory stimuli but may describe a more general organizational principle that applies to all information processing in the brain.

An alternative explanation (less interesting from a cognitive or neural viewpoint) is that the observed spatial pattern is a result of local changes in cerebral blood flow or volume, possibly reflecting active or passive hemodynamic processes. As blood flow increases in task positive regions, to meet local metabolic needs, neighboring regions experience compensatory reductions in blood flow ("vascular steal"), leading to local changes in BOLD (25, 26). However, while vascular stealing has not been definitely ruled out as an explanation of findings of negative BOLD, a number of studies have robustly shown that negative BOLD reflects underlying neural activity (27). Further, we observe the spatial relationship even when only considering task positive and task negative voxels that are a minimum distance (>8mm) apart (a distance more than double most investigations of underlying BOLD point-spread functions (e,.g., (28). Therefore, it is unlikely that our results can be explained as a purely hemodynamic phenomenon.

Here, we have provided evidence for some form of spatial relationship operating between task positive and task negative networks. More work is needed to explore how this relationship operates across scales, e.g., e.g. not just between two generic systems (i.e. the TP and TN networks) but also within networks and lower still.). Equally, further work is needed to investigate different methods for measuring spatial differences between task-evoked patterns of activation and deactivation across different tasks. Here, we used Euclidean distances in three-dimensional MNI space since we wanted to allow for both small (intraregional) distances as well as interregional distances. When looking within a region, distance measured on the two-dimensional unfolded cortical surface may be more appropriate. A different approach for assessing the existence and character of the TP-TN spatial relationship would use diffusion imaging to assess if the long-range white-matter tract organization reflects these spatial patterns.